\documentclass[10pt,aps,prx,twocolumn,notitlepage,showpacs,superscriptaddress, 
]{revtex4-1}
\usepackage{amsthm,amsmath,amsopn,amsfonts,amssymb,verbatim,color}
\usepackage{graphicx}
\usepackage{subfigure}
\usepackage{bm}
\usepackage{epsfig,slashed}
\usepackage[T1]{fontenc}
\usepackage[colorlinks=true,citecolor=blue,linkcolor=blue,urlcolor=blue]{hyperref}
\usepackage{psfrag}

\usepackage{color} 

\begin{document}

\newcommand{\tr}{\mathop{\mathrm{Tr}}}
\newcommand{\bsigma}{\boldsymbol{\sigma}}
\newcommand{\re}{\mathop{\mathrm{Re}}}
\newcommand{\im}{\mathop{\mathrm{Im}}}
\renewcommand{\b}[1]{{\boldsymbol{#1}}}
\newcommand{\diag}{\mathrm{diag}}
\newcommand{\sign}{\mathrm{sign}}
\newcommand{\sgn}{\mathop{\mathrm{sgn}}}
\renewcommand{\c}[1]{\mathcal{#1}}
\renewcommand{\d}{\text{\dj}}
\newcommand{\red}{\textcolor{red}}
\newcommand{\green}{\textcolor{green}}

\newcommand{\mb}{\bm}
\newcommand{\ua}{\uparrow}
\newcommand{\da}{\downarrow}
\newcommand{\ra}{\rightarrow}
\newcommand{\la}{\leftarrow}
\newcommand{\mc}{\mathcal}
\newcommand{\bs}{\boldsymbol}
\newcommand{\lra}{\leftrightarrow}
\newcommand{\nn}{\nonumber}
\newcommand{\half}{{\textstyle{\frac{1}{2}}}}
\newcommand{\mf}{\mathfrak}
\newcommand{\MF}{\text{MF}}
\newcommand{\IR}{\text{IR}}
\newcommand{\UV}{\text{UV}}
\newcommand{\be}{\begin{equation}}
\newcommand{\ee}{\end{equation}}
\newcommand{\eref}{\eqref}

\title[Mixed temperature-dependent order parameters in the extended Hubbard model]{Mixed temperature-dependent order parameters in the extended Hubbard model}

\author{Joel Hutchinson}
\email[electronic address: ]{joel.hutchinson@polytechnique.edu}
\affiliation{Department of Physics, University of Alberta, Edmonton, Alberta T6G 2E1, Canada}
\affiliation{Centre de physique th{\' e}orique, {\' E}cole Polytechnique, 91120 Palaiseau, France}

\author{Frank Marsiglio}
\email[electronic address: ]{fm3@ualberta.ca}
\affiliation{Department of Physics, University of Alberta, Edmonton, Alberta T6G 2E1, Canada}
\affiliation{Theoretical Physics Institute, University of Alberta, Edmonton, Alberta T6G 2E1, Canada}

\vspace{10pt}

\begin{abstract}
The extended Hubbard model can host $s$-wave, $d$-wave and $p$-wave superconducting phases depending on the values of the on-site and nearest-neighbour interactions. Upon detailed examination of the free energy functional of the gap in this model, we show that these symmetries are often dependent on temperature. The critical points of this functional are constrained by symmetry and allow us to formulate stringent conditions on the temperature profile of the gap function, applicable to other models as well. We discuss the finite temperature phase diagram of the extended Hubbard model, and point out the existence of symmetry transitions below $T_c$. Understanding the nature of these transitions is crucial to assessing the symmetry of unconventional superconductors.
\end{abstract}

\maketitle
%
%
%
%
%

\section{Introduction}
As more and more unconventional superconductors are discovered, the question of the symmetry of the superconducting order parameter has moved to the forefront as one of the most immediate and important questions to answer about any new material, especially given its close connection to the (often unknown) pairing mechanism. Experimentally, this question is difficult to answer, as the arduous history of the cuprates provides testament for~\cite{annett1996}. The resolution of this question has been aided, in part, by phase-sensitive tunnelling measurements, which have been particularly useful in uncovering the gap symmetry of the heavy-fermion compound UPt$_3$~\cite{strand2009}, shown to have different symmetries of the order parameter over different temperature ranges~\cite{norman2011}. Similarly, recent observations in LaAlO$_3$/SrTiO$_3$~\cite{stornaiuolo2017} and underdoped La$_{2-x}$Sr$_x$CuO$_4$ (LSCO)~\cite{razzoli2013} suggest that a second component of the gap function develops below $T_c$. 
The purpose of this paper is to point out that one should generically expect to find a rich phase diagram below $T_c$, particularly at strong coupling. In fact we illustrate that symmetry transitions occur as a function of temperature in one of the simplest and most studied models for superconductivity, the extended Hubbard model on a square lattice. More particularly, this occurs in the regime of attractive nearest-neighbour interactions ($V<0$).

In one dimension (1D), the phase diagram of the extended Hubbard model with $V<0$ has been well established. In the presence of on-site attraction ($U<0$), the system hosts singlet superconductivity, while above a critical $U$, this transitions to triplet superconductivity; spin-density-wave order (SDW) and phase separation are observed as well~\cite{kuroki1994, zhi2002, sengupta2002}. 
A definitive phase diagram in two dimensions (2D) is more difficult to produce owing in part to the presence of a sign problem~\cite{santos2003}. On a bipartite lattice such as the square lattice, sign-free quantum Monte Carlo away from half-filling requires attractive on-site, and repulsive nearest-neighbour interactions~\cite{wei2016}. As a result, the $V<0$ regime is largely unexplored. This is unfortunate because an attractive inter-site interaction can arise, for example, due to polarons, as in the iron-pnictides~\cite{sawatzky2009}. Thus, there are only a few benchmarks we use to frame the phase diagram of this region within. The ($V=0,\; U<0$) region is dominated by $s$-wave superconductivity~\cite{singer1998}, and at half filling, the $U>0$ line contains an SDW transition~\cite{wu2014, buividovich2018}. Doping slightly away from half filling yields $d$-wave superconductivity according to dynamical mean field theory~\cite{jarrell2001, civelli2008}, which makes this a good candidate model for the high-$T_c$ cuprates. 

If we are to take any lessons from the exact results in 1D, it should be that both singlet and triplet superconductivity are present in the extended Hubbard model. However, in 2D we have the additional angular freedom of the gap profile in momentum-space. This makes the list of possible angular profiles much larger. In fact, as we show in this paper, one should not generically expect any rotation symmetries of the lattice to be retained by the gap function well below $T_c$. 

Even when numerically exact methods like quantum Monte Carlo can be used, the question of the symmetry of the gap is far from answered. Such methods look for instabilities in specific angular channels and can easily miss mixed phases. Indeed as we will see, the energetic difference between mixed and non-mixed phases can be very small. In this paper we provide a systematic approach to find the exact profile of the gap function in momentum-space within the context of mean-field theory. 

Constraints on the symmetry of the gap function are partially established within Landau-Ginzburg theory~\cite{annett1990}. In this paradigm, the gap function is segmented into pieces that transform under irreducible representations of the normal state symmetry group. Any of these individual pieces could form the superconducting state at $T_c$, but cannot be mixed at this temperature due to invariance of the free energy. However, at lower temperatures, when the magnitude of the order parameter is large, higher order terms in the Landau free energy become important. At these lower temperatures, the free energy becomes a complicated function of the gap, so invariance of the free energy under lattice symmetries does not impose such stringent conditions on the gap. As a result, mixing can occur. As we will see, the phases below $T_c$ are simply described by pitchfork bifurcations of critical points of the free energy. Such mixing and bifurcations have been predicted before in the context of anisotropic tight binding models~\cite{sorensen1991, angilella1999}. In this paper, we illustrate these ideas within a case study of the 2D extended Hubbard model. 
The zero-temperature superconducting phase diagram for this model is quite rich~\cite{nayak2018}. 
We provide an efficient numerical method for determining the mean-field phase diagram, that is able to capture a phase that was missed with previous techniques. Our results indicate that many superconductors described by the extended Hubbard model are likely not strictly ``s-wave'' or ``d-wave'' but instead consist of gap functions that acquire multiple components below $T_c$.

\section{Model and Mean Field Solution}
We consider the extended Hubbard model on a square lattice with unit lattice spacing, nearest neighbour hopping $t$, on-site interaction $U$, nearest-neighbour interaction $V$, and chemical potential $\mu$
\begin{eqnarray}\label{eq:H1}
H &=&-t\sum_{\langle i,j \rangle \atop \sigma} (c_{i\sigma}^{\dagger
}c_{j\sigma}+c_{j\sigma}^{\dagger }c_{i\sigma})  + U\sum_i n_{i\uparrow} n_{i\downarrow}\nonumber\\
&& + V \sum_{\langle  ij \rangle \atop \sigma , \sigma^\prime} n_{i\sigma} n_{j\sigma^\prime}-\mu\sum_{i,\sigma}n_{i\sigma},
\end{eqnarray}
where $\langle i, j\rangle$ denotes nearest neighbour sites and $\sigma$ is a spin index. 
The standard pairing interaction, treated in all spin channels, results in a $2\times2$ matrix $\Delta_{\alpha\beta}$, whose components satisfy a set of self-consistent gap equations (see appendix~\ref{app:mft} for details). Since the purpose of this paper is to differentiate the competing superconducting symmetries, we focus on the attractive nearest-neighbour regime ($V<0$). In the half-filled positive $U$ regime, there is competing spin-density wave order with higher $T_c$~\cite{micnas1990, robaszkiewicz1981}, but renormalization group results at zero temperature indicate that this phase does not survive much below $n=1$~\cite{huang2013}. Based on this evidence it seems unlikely that charge or spin-density waves play a significant role for $V<0, n<1$. We can therefore take our mean fields to include only superconducting order parameters $\Delta_{\alpha\beta}$.

\begin{figure}[h]
	\centering
	\includegraphics[width=0.9\columnwidth]{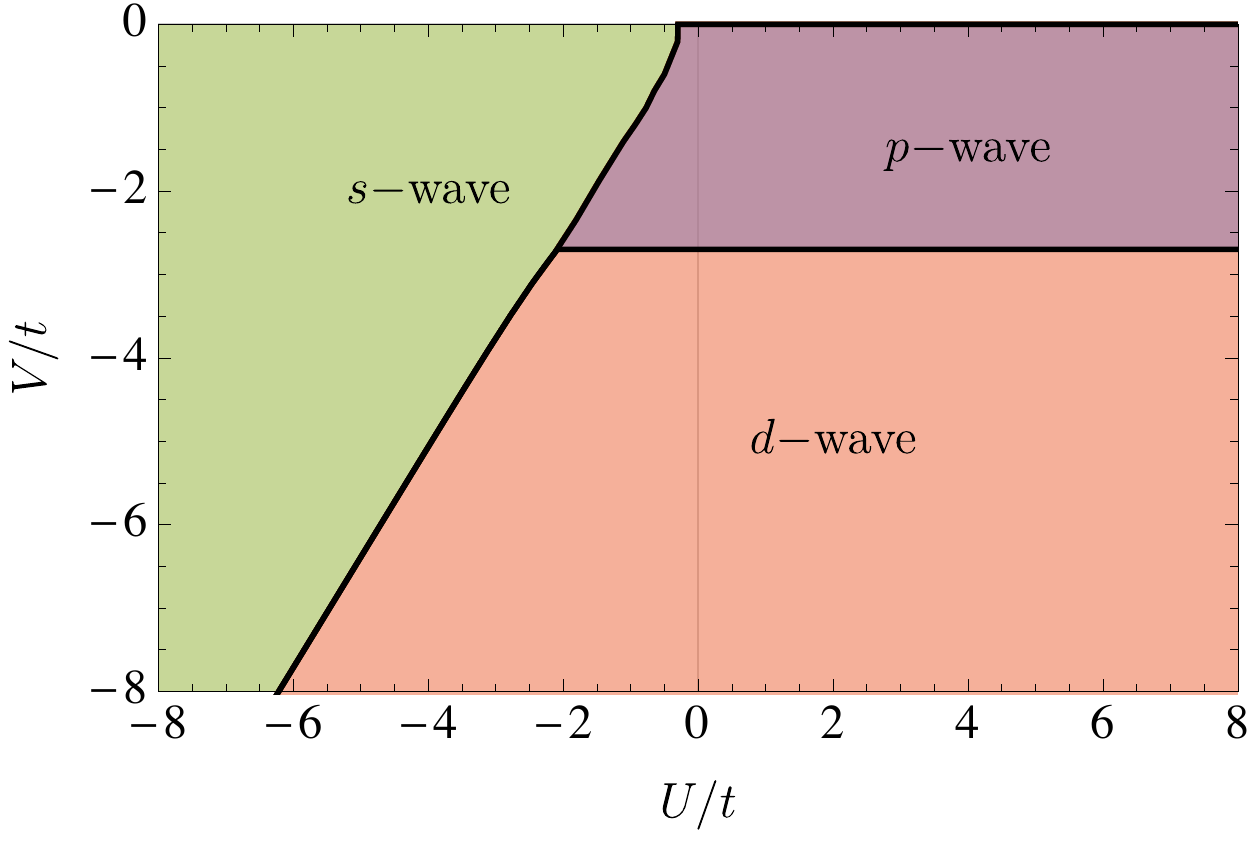}
\caption{Phase diagram based on $T_c$ for electron density $n=0.5$. Black lines highlight the phase boundaries. 
}\label{fig:tcplt}
\end{figure}

First we consider the situation near $T_c$ where all gap components are small. The phase diagram at $T_c$ has been studied before~\cite{micnas1990}. The gap equations decouple into three different sets of equations that correspond to the $s$-wave, $d$-wave and triplet $p$-wave phases. 
In general, each irreducible representation of the lattice point group will have a corresponding $T_c$ equation provided there is a part of the interaction that transforms under this representation. Unlike the $d$-wave case, the linearized $s$-wave gap equations can have two eigenvalues in some parts of the parameter space and therefore two critical temperatures.  In Fig.~\ref{fig:tcplt}, we show the phase diagram in the $U$-$V$ plane at quarter filling based solely on the critical temperatures determined by solving the linearized gap equations. Often, this diagram is taken to give the symmetry of the order parameter below $T_c$. However, we will see that the true phase diagram is much richer when we take into account the temperature dependence of the gap.

For the sake of generality, we may rewrite the gap function as
\be\label{eq:Deltagen}
\Delta^{\rm sing}_\b{k}=\sum_{i\in\{0,s,d\}}\Delta_ix^i_\b{k};\;\;\;\; \Delta^{\rm trip}_\b{k}=\sum_{i\in\{x,y\}}\Delta^{\rm trip}_ix^i_\b{k}.
\ee
Here the basis functions $x^i_\b{k}$ are 
\begin{eqnarray}
x^0_\b{k}&=&1,\label{eq:basis1}\\ 
x^s_\b{k}&=&s_\b{k},\\ 
x^d_\b{k}&=&d_\b{k},\\
 x^x_\b{k}&=&\sin k_x,\\
x^y_\b{k}&=&\sin k_y,\label{eq:basislast}
\end{eqnarray}
where $s_\b{k}\equiv\frac{1}{2}(\cos k_x+\cos k_y)$ and $d_\b{k}\equiv\frac{1}{2}(\cos k_x-\cos k_y)$. Such a decomposition can be made for any separable interaction and makes the expressions we derive valid for many models beyond the extended Hubbard model. Note that $x^0_\b{k}$ and $x^s_\b{k}$ belong to one irreducible representation of the square lattice point group ($A_1$), while $x^d_\b{k}$ belongs to  $B_1$ and ($x^x_\b{k}$, $x^y_\b{k}$) belong to another ($E$). For the singlet part, this means that the gap equations decouple whenever $\Delta_d=0$ or $\Delta_0=\Delta_s=0$. We refer to such solutions as pure solutions, and reserve ``mixed solutions'' for any case where components from multiple 
{\it different} irreducible representations are non-zero. 

The natural question to ask is which symmetry of the gap would be observed for a given value of the interactions $U$, $V$, temperature $T$ 
and electron density $n$? The answer is whatever minimizes the free energy density. Minimization of the free energy is an important problem that requires strict numerical control, and we have developed a new method to deal it.

In appendix~\ref{app:free} and~\ref{app:critical}, we show that the global minimum of the free energy proceeds through a series of bifurcations in parameter space as the temperature is lowered. Knowledge of these bifurcations provides an important numerical advantage. Identification of the symmetry of the order parameter at a given temperature requires the global minimization of a multidimensional function that is quite nonlinear. Without making use of these bifurcations, one would have to proceed by brute force search of a discretized parameter space in order to find all local minima~\cite{nayak2018}. Instead, we leverage the bifurcations by doing a minimization within the lower-dimensional pure subspaces of the parameter space. The set of local minima found in these subspaces might not contain the global minimum, but one of them will lie within the basin of attraction of the global minimum. We can therefore initiate a local minimization scheme at each of these points and one of them will converge to the global minimum.

This procedure provides a significant reduction in computational time, scaling as the dimension of the largest pure subspace (2 in this model) rather than the dimension of the full paramter space (5 in this model). It may also be applied to any free energy minimization problem in which the critical points are continuous functions of some parameter (in this case temperature). For a single order parameter, this is equivalent to saying that the system only has a second-order phase transition. However, this does not preclude the existence of first-order transitions in the case of multiple competing components of the order parameter as we will see in Sec.\ref{sec:phase}.

\section{Phase diagram}\label{sec:phase}

\begin{figure}[t]
	\centering
	\includegraphics[width=0.88\columnwidth]{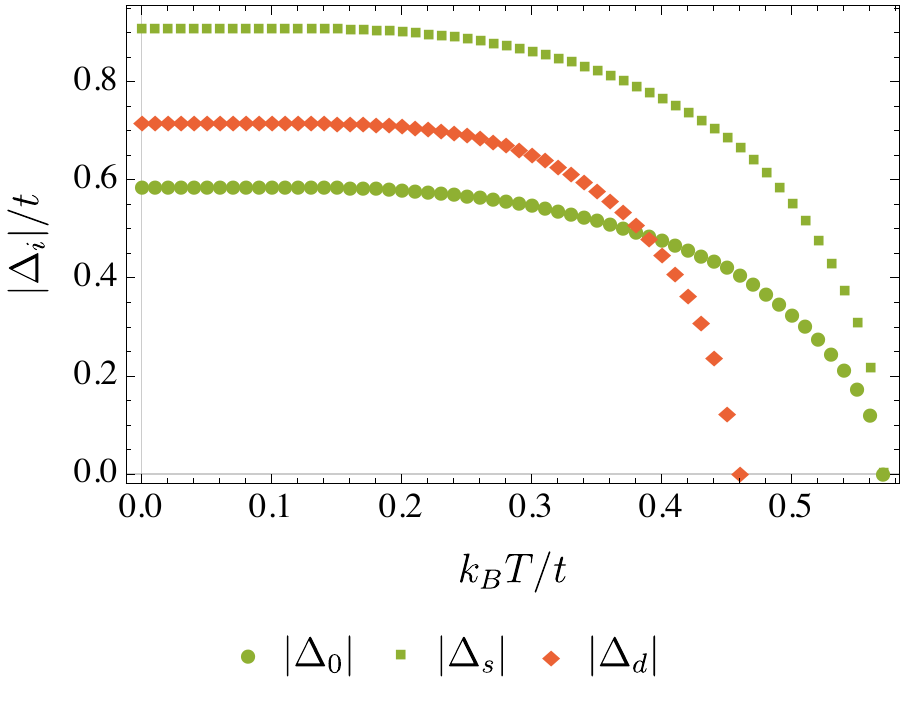}
\caption{(a) Temperature dependence of the gap components at $U=-3.0$, $V=-3.3$, $n=0.5$. The $s$-wave part (green) has two components ($\Delta_0$ and $\Delta_s$). The red dots show $|\Delta_d|/t$, which emerges at a lower temperature and comes with a relative phase of $\pi/2$ with respect to the $s$-wave components, forming an emergent $s+id$ phase.}\label{fig:temp_prof1}
\end{figure}

An example of a point in the phase diagram where a mixed solution takes over below $T_c$ is shown in Fig.~\ref{fig:temp_prof1}. For these parameters, the system starts as a $s$-wave superconductor at $T_c^s$, but attains a lower free energy upon the emergence of $d$-wave components at a lower temperature forming an $s+id$ phase. 

\begin{figure}[t]
	\centering
	\includegraphics[width=0.88\columnwidth]{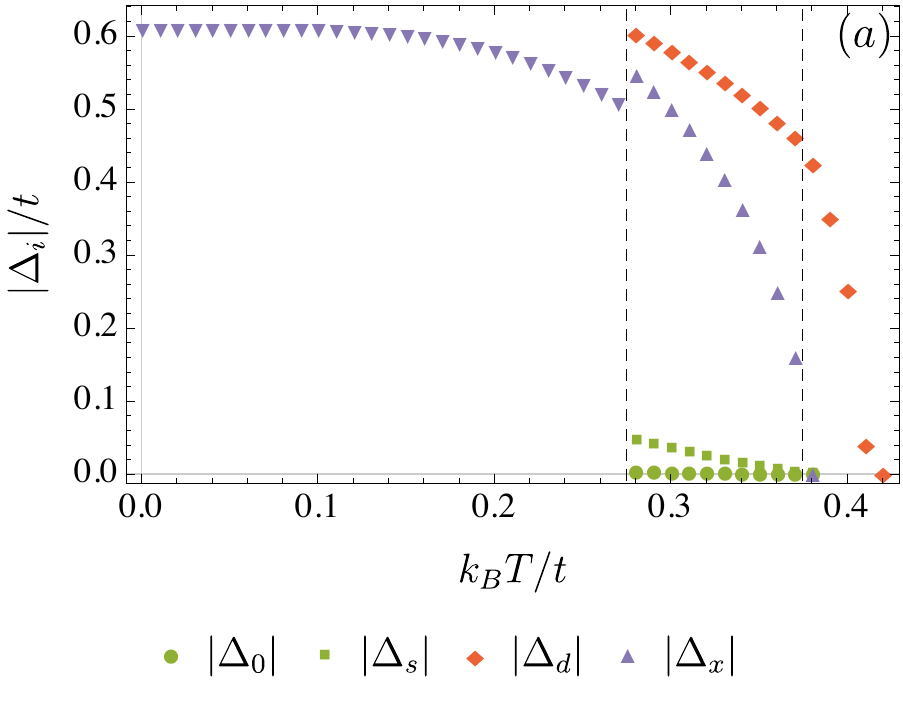}\\
	\includegraphics[width=0.88\columnwidth]{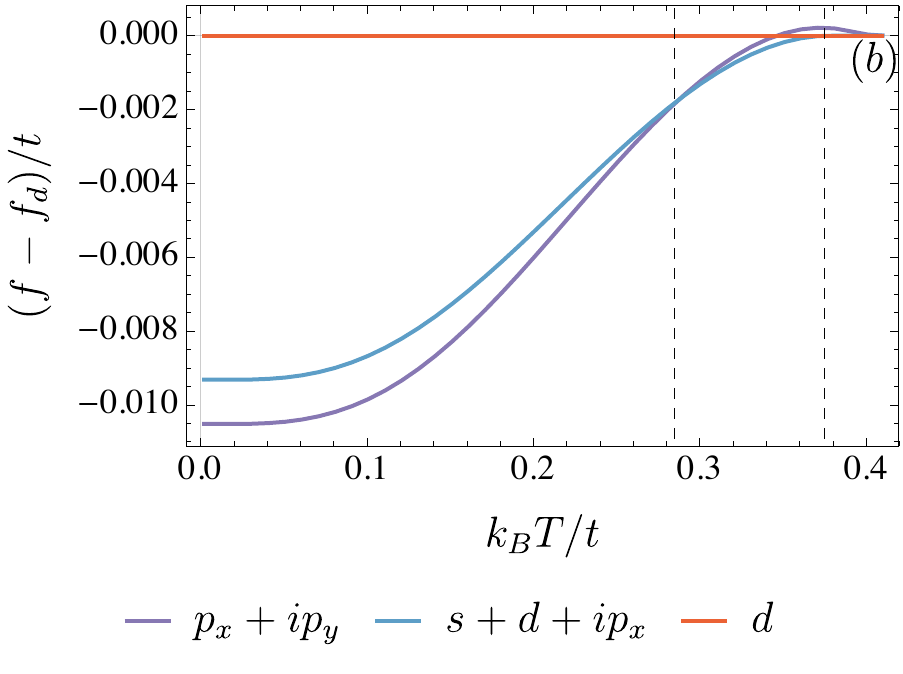}
\caption{(a) Temperature dependence of the gap components for the global minimum at $U=-0.5$, $V=-3.05$, $n=0.5$. A transition occurs from $d$-wave to the mixed state $s+d+ip_x$ (which is degenerate with  $s+d+ip_y$) at $k_{\rm B} T=0.375t$, and a second transition occurs at $k_{\rm B} T=0.285t$ (indicated by the dashed lines). (b) Free energy density of the candidate minima relative to the free energy density of the $d$-wave solution, showing how the pure $p+ip$ solution overtakes the others at low temperatures.}\label{fig:temp_prof2}
\end{figure}

Fig.~\ref{fig:temp_prof2}(a) shows a more complicated situation that is prevalent near the $T_c^d=T_c^p$ line. Here we see the development of a free energy minimum that mixes all symmetry sectors $s$-wave, $d$-wave and $p$-wave at a second-order transition below $T_c$. At an even lower temperature, this minimum gives way to the pure $p$-wave solution. Remarkably, both the $\b{k}$-space symmetry and spin pairing change twice within the superconducting phase. 
Note that even the $p$-wave component is not continuous across this latter transition because the solutions on either side of the transition descend from different bifurcations of the free energy critical points. The $p$-wave component above the transition emerges from the $d$-wave phase, while the one below the transition emerges from the normal state in accordance with the bifurcation rules discussed in appendix~\ref{app:critical}. This is more easily seen by comparing the free energies as in Fig.~\ref{fig:temp_prof2}(b). Temperature profiles such as this would be accompanied by a jump in the specific heat at the onset of the mixed phase, as well as a second anomaly at lower temperature due to the reappearance of a pure phase. However, we note that the derivative of the free energy across the latter transition is remarkably smooth, as seen in Fig.~\ref{fig:temp_prof2}(b), so the corresponding anomaly may be difficult to observe. 

\begin{figure}
	\centering
	\includegraphics[width=0.99\linewidth]{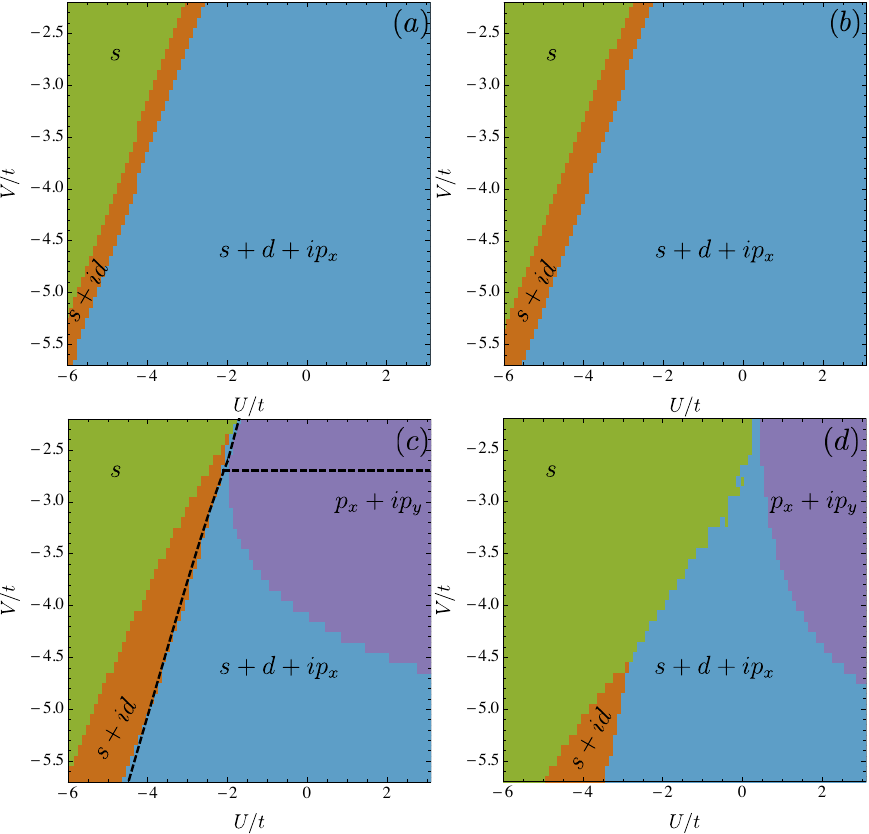}
\caption{Numerically determined zero temperature $U$-$V$ phase diagram for densities $n=0.9$ (a), $n=0.7$ (b), $n=0.5$ (c), $n=0.3$ (d). The dashed lines on (c) indicate the phase boundaries at $T_c$ corresponding to Fig.~\ref{fig:tcplt}.}
\label{fig:phase_diag}
\end{figure}

The zero-temperature phase diagrams are shown in Fig.~\ref{fig:phase_diag} for different densities. Near half-filling, the phase diagram is dominated by a large mixed phase containing $s$, $p$ and $d$-wave components. Within this phase, the $s$ and $p$-wave components decrease as $V$ increases towards zero, first approaching a $d+ip_x$ phase, followed by a pure $d$-wave phase at weak coupling (not shown here). At lower fillings, a $p+ip$ phase is stabilized in agreement with the $T_c$ results, and an $s+id$ phase exists only near the border between $s$ and $s+d+ip_x$ phases, for all fillings. In Fig.~\ref{fig:phase_diag} (c), we overlay the $T_c$ phase boundaries at quarter filling. This illustrates the dramatic distinction between the high and low temperature superconducting phases. The large region of the $d$-wave phase near $T_c$ is completely replaced at low temperatures by a phase with more broken symmetries. Even near half-filling, where the extended Hubbard model serves as a paradigmatic model for high-$T_c$ $d$-wave superconductors, the $d$-wave phase is mostly replaced by a mixed phase at low temperatures. Moreover, this rules out the possibility that such high-$T_c$ superconductors have a low temperature $s+id$ phase, since this $s+id$ phase occurs entirely to the left of the $T_c^s=T_c^d$ line.

\begin{figure}
	\centering
		\includegraphics[width=0.88\columnwidth]{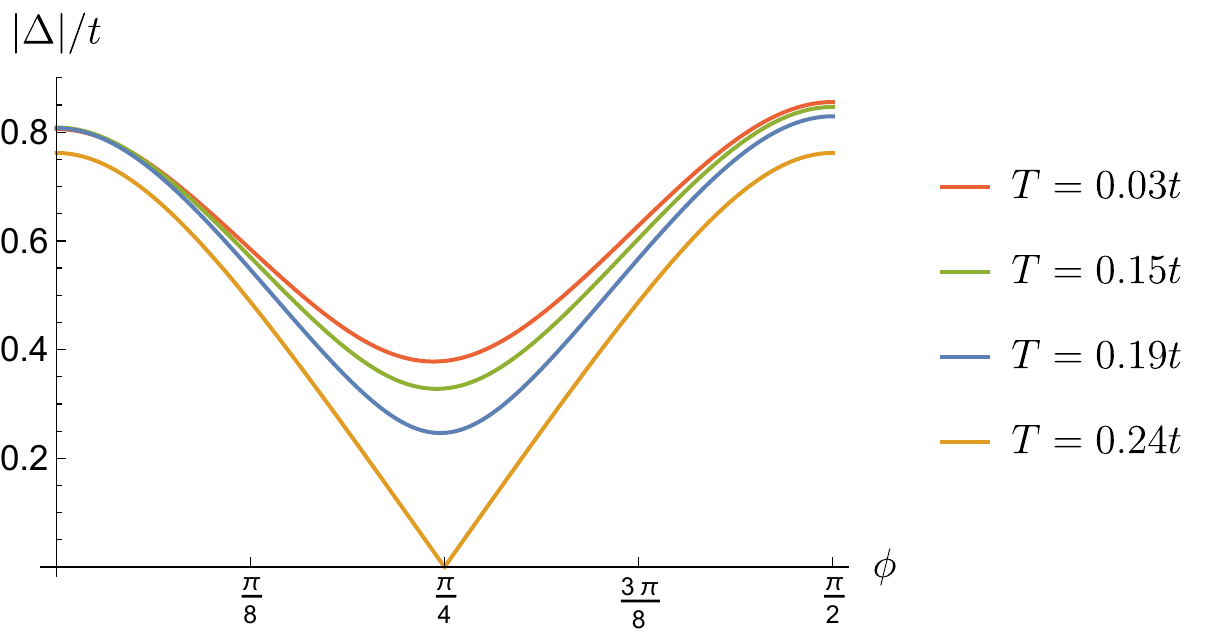}
\caption{Magnitude of the zero-temperature gap as a function of angle $\phi\equiv\arctan(k_y/k_x)$ on the Fermi surface for $n=0.7$, $U=-t$, $V=-2t$, where the gap is predominantly $d+ip_y$. 
}\label{fig:angular}
\end{figure}

\begin{figure}
	\centering
		\includegraphics[width=0.88\columnwidth]{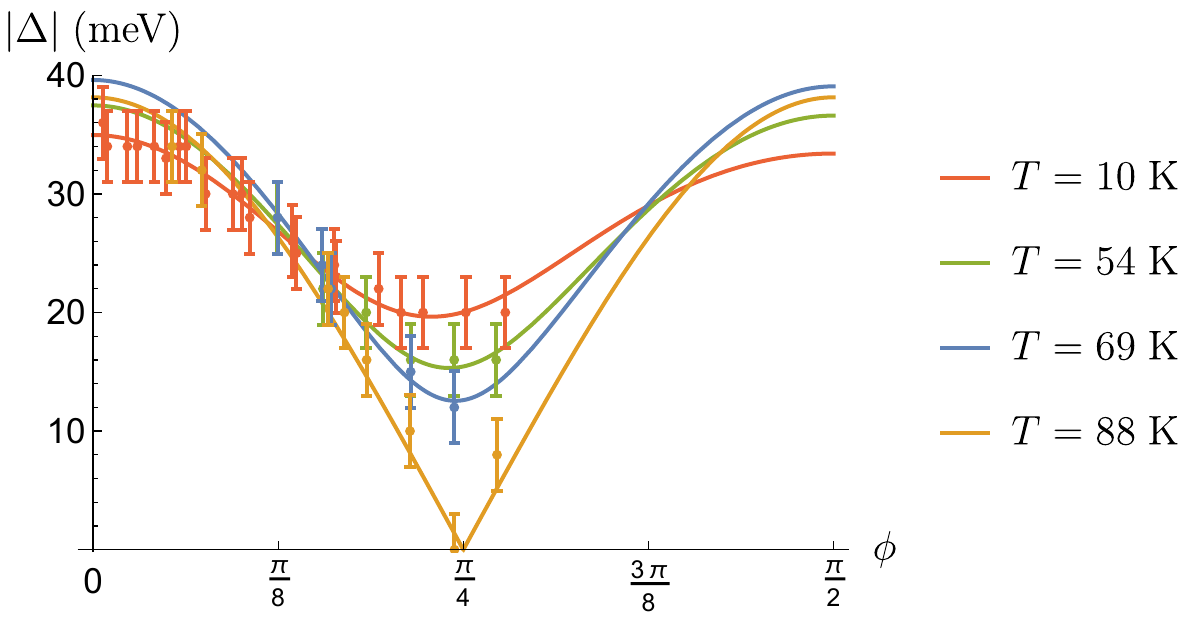}
\caption{Angle-resolved photoemission spectroscopy data of the gap in LSCO with $x=0.08$ taken (with permission) from Razzoli \textit{et al.}~\cite{razzoli2013}. The solid curves show fits to a $d+ip_y$ gap: $\Delta_dd_{\b{k}}+\Delta_p\sin k_y$ where $\Delta_d$ and $\Delta_p$ are fit parameters for each temperature. The gap parameters obtained (in meV) are $\Delta_d=35.7$, $\Delta_p=20.9$ for $T=10$K, $\Delta_d=39.2$, $\Delta_p=15.1$ for $T=54$K, $\Delta_d=41.8$, $\Delta_p=13.1$ for $T=69$K, $\Delta_d=39.2$, and $\Delta_p=15.1$ for $T=54$K, $\Delta_d=40.8$, $\Delta_p=0.0$ for $T=88$K.}\label{fig:experiment}
\end{figure}

Fig.~\ref{fig:angular} shows the angular dependence of the magnitude of the gap in the mixed phase. In this example, the $s$-wave component is very small so that the gap is effectively $d+ip_y$ (which is degenerate with $d+ip_x$). We see that the $p$-wave component opens a gap at the $d$-wave nodes and introduces a slight asymmetry between $\phi=0$ and $\phi=\pi$. In general, mixed gap functions can have zero, two or four nodes depending on the fermi level and the relative magnitude of the $\Delta_i$'s. This figure provides a prediction for photoemission experiments in some cuprate superconductors. While many cuprates display a node for all temperatures (e.g. Bi$_2$Sr$_2$CaCu$_2$O$_{8+\delta}$~\cite{hashimoto2014}), this is not true in the deeply underdoped regime, particularly for underdoped LSCO~\cite{razzoli2013}, which shows similar temperature dependence to Fig.~\ref{fig:angular}. In this case, the authors of Razzoli \textit{et al.}~\cite{razzoli2013} go beyond $T_c$ into the pseudogap phase. However, because mean field theory does not account for phase fluctuations, our results really probe the spectral gap, as discussed in Emery \textit{et al.}~\cite{emery1995}. The nature of the pseudogap is of course widely debated, but we see that the changing symmetry of the energy gap in this region should not be construed as evidence for some competing order. On the other hand, our results show that one should generically expect changes to the gap symmetry at intermediate to strong coupling, arising solely from the pairing order parameter. This is consistent with the idea of preformed pairs, which is supported by recent experiments~\cite{zhou2019, yuli2009}. Fig.~\ref{fig:experiment} shows that a transition from a $d$-wave to a $d+ip$ gap at low temperatures indeed fits well with the experimental data for LSCO. 


With these examples, we see that the full temperature-dependent phase diagram is quite complicated. In general, there is a significant portion of the phase diagram where $d$-wave gives way to a mixed state below $T_c$ which may be followed by a $p+ip$ phase at even lower temperatures.  

\section{Conclusion} 
We have studied the temperature dependence of the symmetry of the superconducting gap in the 2D extended Hubbard model, focusing on the regime of attractive nearest neighbour interaction ($V<0$). We used a mean field approach to determine the gap equations that set the allowed symmetries of the gap.

The precise symmetry of the gap for given parameters ($U,V,n,T$) is determined by the minimum of the free energy, which can be considered as a critical point of a Morse function in a multi-dimensional parameter space. This perspective, outlined in the appendix~\ref{app:free} and~\ref{app:critical}, has conceptual and technical advantages that may translate to other models as well. 

We discovered that even within this simple model, there is a rich variety of symmetry phase transitions that occur as a function of temperature, observable through the specific heat or photoemission. In particular, we note that there are portions of the phase diagram where there is a different pure symmetry at $T=T_c$ and $T=0$ (e.g $d$-wave and $p+ip$), with multiple symmetry transitions at intermediate temperatures. It is interesting that even deep within the parameter range where $d$-wave has the highest $T_c$ (i.e. far from the $T_c$ phase boundaries), where we expect that different symmetries are least competitive, the mixed solution is dominant at low temperatures. At zero temperature where the magnitude of the gap is large, it appears that $d$-wave symmetry is never preferred at strong coupling, and the energy can always be lowered by the addition of $s$ or $p$-wave components. Provided these new phases are stable to quantum fluctuations, there are two possibilities; either the extended Hubbard model is not sufficient to capture the symmetry of $d$-wave superconductors with large $V$, or experiments to determine the gap symmetry should include fits to more complicated gap functions containing small $s$ or $p$-wave components. Indeed, we have compared the temperature-dependence found in the mixed phase with experiments in underdoped LSCO and found a striking similarity. It should be noted that the presence of p-wave in our result produces an asymmetry between the antinodes of the gap that was not tested for experimentally. Of course, it is still possible that the symmetry is given by $d_{x^2-y^2}+d_{xy}$, as suggested in Razzoli \textit{et al.}~\cite{razzoli2013}. More low-temperature data at both antinodes is needed to differentiate these two possibilities. $d_{xy}$ components do not appear in the extended Hubbard model, but may be included with e.g. a next-nearest neighbour hopping term. In that case one might find transitions from pure $d_{x^2-y^2}$ to mixed $d_{x^2-y^2}+d_{xy}$.

A word about the interaction parameter ranges in this paper is warranted. We focused mainly on strong nearest-neighbour coupling, and found that in this regime, the range of symmetry transitions is significant. Such transitions may also occur at weak coupling, but in this regime the free energies of the different phases are so similar as to make it difficult to discern the transition. Therefore it would be easiest to observe this phenomenon for large $|V|/t$. One promising avenue to explore this is in the context of ultra-cold atoms, where $U/t$ is controlled via the optical lattice potential, $n$ and $T$ are controlled via evaporative cooling, and $V$ can be included through dipolar interactions~\cite{tarruell2018, menotti2008}. 

Lastly, we point out that the phase diagram is highly dependent on the symmetries of the normal state and therefore may be further enriched by the addition of Rashba or Zeeman terms.

\acknowledgements
We wish to thank Elia Razzoli for sharing photoemission data with us. 
This work was supported in part
by the Natural Sciences and Engineering Research Council of
Canada (NSERC) as well as Alberta Innovates - Technology Futures (AITF). 
J.H. was supported by fellowships from NSERC, AITF, and the French National Research Agency (ANR).

\appendix

\section{Mean Field Theory}\label{app:mft}
After Fourier transforming the Hamiltonian we restrict our consideration to interactions between pairs of zero total momentum; the absence of finite-momentum pairing is expected in an unpolarized system. The resulting Hamiltonian reads
\begin{eqnarray}
H&=&\sum_{\mathbf{k},\sigma}\xi_\mathbf{k}c^\dagger_{\mathbf{k}\sigma}c_{\mathbf{k}\sigma}\nonumber\\
&&+\frac{1}{2N}\sum_{\b{kk}'}\sum_{\alpha\beta\gamma\delta}V_{\alpha\beta\gamma\delta}(\b{k},\b{k}')c^\dagger_{\b{k}\alpha} c^\dagger_{-\b{k}\beta}c_{-\b{k}'\gamma}c_{\b{k}'\delta}.\nonumber\\\label{eq:Hk}
\end{eqnarray}
Here $\xi_\mathbf{k}=-2t(\cos k_x +\cos k_y)-\mu$, $N$ is the total number of sites in the lattice and the non-zero components of the interaction are
\begin{eqnarray}
V_{\uparrow\downarrow\downarrow\uparrow}=V_{\downarrow\uparrow\uparrow\downarrow}&=&\frac{1}{2}[U+4V(s_\b{k}s_{\b{k}'}+d_\b{k}d_{\b{k}'})\nonumber\\
&&+2V(\sin k_x\sin k_x'+\sin k_y\sin k_y')],\nonumber\\
\\
V_{\uparrow\downarrow\uparrow\downarrow}=V_{\downarrow\uparrow\downarrow\uparrow}&=&\frac{1}{2}[-U-4V(s_\b{k}s_{\b{k}'}+d_\b{k}d_{\b{k}'})\nonumber\\
&&+2V(\sin k_x\sin k_x'+\sin k_y\sin k_y')],\nonumber\\
\\
V_{\uparrow\uparrow\uparrow\uparrow}=V_{\downarrow\downarrow\downarrow\downarrow}&=&4V(s_\b{k}s_{\b{k}'}+d_\b{k}d_{\b{k}'})\nonumber\\
&&+2V(\sin k_x\sin k_x'+\sin k_y\sin k_y')],\nonumber\\
\end{eqnarray}
where we have used the $s$-wave and $d$-wave basis functions: $s_\b{k}\equiv\frac{1}{2}(\cos k_x+\cos k_y)$, and $d_\b{k}\equiv\frac{1}{2}(\cos k_x-\cos k_y)$.

We apply mean-field theory to Eq.~\eref{eq:Hk} following Ref.~\cite{sigrist2005}. Including all anomalous pairing mean fields \footnote{We ignore Hartree-Fock terms here. These terms will renormalize the bandwidth and the chemical potential, but will not qualitatively change our results~\cite{micnas1990}.},  the generic four-fermion interaction (suppressing momentum labels for now) becomes
\be
c^\dagger_\nu c^\dagger_\mu c_{\mu'}c_{\nu'}\approx\langle c^\dagger_\nu c^\dagger_\mu\rangle c_{\mu'}c_{\nu'}+ c^\dagger_\nu c^\dagger_\mu \langle c_{\mu'}c_{\nu'}\rangle - \langle c^\dagger_\nu c^\dagger_\mu\rangle \langle c_{\mu'}c_{\nu'}\rangle,
\ee
where we have removed terms of order $(c_\sigma c_{\sigma'}- \langle c_\sigma c_{\sigma'}\rangle)^2$ and their hermitian conjugates.

The result is a Hamiltonian with only fermion bilinears:
\begin{eqnarray}\label{eq:HMF}
H_{\rm MF}&=&\sum_{\mathbf{k},\sigma}\xi_\mathbf{k}c^\dagger_{\mathbf{k}\sigma}c_{\mathbf{k}\sigma}-\frac{1}{2}\sum_{\b{k}\sigma\sigma'}\Delta_{\b{k}\sigma\sigma'}c^\dagger_{\b{k}\sigma}c^\dagger_{-\b{k}\sigma'}\nonumber\\
&&-\frac{1}{2}\sum_{\b{k}\sigma\sigma'}\Delta^*_{\b{k}\sigma\sigma'}c_{\b{k}\sigma}c_{-\b{k}\sigma'}+E_{\rm MF},
\end{eqnarray}
where we have defined the gap function
\be\label{eq:gap_gen_eqn}
\Delta_{\b{k}\sigma\sigma'}\equiv-\frac{1}{N}\sum_{\b{k}'\tau\tau'}V_{\sigma\sigma'\tau\tau'}(\b{k},\b{k}')\langle c_{-\b{k}'\tau}c_{\b{k}'\tau'}\rangle,
\ee
and the energy due to the product of mean fields
\be
E_{\rm MF}\equiv\frac{-1}{2N}\sum_{\b{k}\b{k}'}\sum_{\sigma\sigma'\tau\tau'}V_{\sigma\sigma'\tau\tau'}(\b{k},\b{k}')\langle c^\dagger_{\b{k}\sigma}c^\dagger_{-\b{k}\sigma'}\rangle\langle c_{-\b{k}'\tau}c_{\b{k}'\tau'}\rangle.
\ee

This may be compactly written as
\be
H_{\rm MF}=\sum_{\b{k}}(\psi^\dagger_{\b{k}}h_{\b{k}}\psi_{\b{k}}+\xi_{\b{k}})+E_{\rm MF},
\ee
where 
\be
\psi_{\b{k}}\equiv
\begin{pmatrix}
c_{\b{k}\uparrow}\\
c_{\b{k}\downarrow}\\
c^\dagger_{-\b{k}\uparrow}\\
c^\dagger_{-\b{k}\downarrow}
\end{pmatrix}
;\;\;h_{\b{k}}=\frac{1}{2}\begin{pmatrix}
\xi_\b{k}\mathbb{I} & \Delta_\b{k}\\
\Delta^\dagger_\b{k} & -\xi_\b{k}\mathbb{I}
\end{pmatrix}.
\ee
Here $\mathbb{I}$ is the $2\times2$ identity matrix, and $\Delta_\b{k}$ is the $2\times2$ matrix with components $\Delta_{\b{k}'\sigma\sigma'}$. We will exclusively consider gap functions that are unitary: $\Delta^\dagger_\b{k}\Delta_\b{k}=|\Delta_\b{k}|^2\mathbb{I}$, where $|\Delta_\b{k}|^2\equiv\frac{1}{2}\tr\Delta^\dagger_\b{k}\Delta_\b{k}$. A non-unitary gap allows for spin-polarized mean fields ($\Delta_{\uparrow\uparrow}$ and $\Delta_{\downarrow\downarrow}$) to be arbitrary, and are only warranted in proximity to a ferromagnetic instability~\cite{sigrist2005}. In particular, all superconductors that preserve time-reversal symmetry are unitary, as seen by the operation of time-reversal on the gap function $\Delta_\b{k}\rightarrow\sigma_y\Delta^*_\b{k}\sigma_y$. Moreover, some time-reversal symmetry-broken phases are unitary, including the $s+id$ phase.

$h_{\b{k}}$ is diagonalized by the Bogoliubov transformation
where
\begin{eqnarray}
u_\b{k}&\equiv&\frac{E_\b{k}+\xi_{\b{k}}}{\sqrt{2E_\b{k}(E_\b{k}+\xi_\b{k})}}\mathbb{I},\\
v_\b{k}&\equiv&\frac{-1}{\sqrt{2E_\b{k}(E_\b{k}+\xi_\b{k})}}\Delta_{\b{k}},\\
E_\b{k}&\equiv&\sqrt{\xi_\b{k}^2+|\Delta_\b{k}|^2}.
\end{eqnarray}
Here we have used the fact that antisymmetry of $\Delta_\b{k}$ requires that $v^T_{-\b{k}}=-v_\b{k}$ and $v^*_{-\b{k}}=-v^\dagger_\b{k}$ and we have defined the quasi-particle spinors 
\be
\Gamma_\b{k}=\begin{pmatrix}
\gamma_{\b{k}\uparrow}\\
\gamma_{\b{k}\downarrow}\\
\gamma^\dagger_{-\b{k}\uparrow}\\
\gamma^\dagger_{-\b{k}\downarrow}
\end{pmatrix}
\ee 
via $\psi_\b{k}=U_\b{k}\Gamma_\b{k}$. We have also identified the ground state energy
\be
E_g=E_{\rm MF}+\sum_{\b{k}}(\xi_\b{k}-E_\b{k}).\label{eq:Eg_extend}
\ee
In terms of these quasi-particle operators, we have
\begin{eqnarray}
c_{\b{k}\sigma}&=&u_{\b{k}\sigma\sigma}\gamma_{\b{k}\sigma}+\sum_{\sigma'}v_{\b{k}\sigma\sigma'}\gamma^\dagger_{-\b{k}\sigma'}\label{eq:bogo_extend1}\\
c_{-\b{k}\sigma}&=&u_{\b{k}\sigma\sigma}\gamma_{-\b{k}\sigma}-\sum_{\sigma'}v_{\b{k}\sigma'\sigma}\gamma^\dagger_{\b{k}\sigma'}.\label{eq:bogo_extend2}
\end{eqnarray}
This allows us to evaluate the mean-field expectation value $\langle c_{-\b{k}\tau}c_{\b{k}\tau}\rangle$ since the $\gamma_\b{k}$ operators satisfy the fermion anticommutation relations and therefore have the occupation number distribution $\langle \gamma^\dagger_{\b{k}\sigma}\gamma_{\b{k}\sigma}\rangle=f(E_\b{k})$, where $f(E)$ is the Fermi function. The defining gap equation \eref{eq:gap_gen_eqn} then becomes the self-consistency condition
\be
\Delta_{\b{k}\sigma\sigma'}=-\frac{1}{N}\sum_{\b{k}\tau\tau'}V_{\sigma\sigma'\tau\tau'}(\b{k},\b{k}')\Delta_{\b{k}'\tau'\tau}g_{\b{k}'},
\ee
where we have defined  $g_{\b{k}}\equiv\frac{1}{2E_\b{k}}(1-2f(E_{\b{k}}))$.

It is conventional to separate the gap matrix into its singlet and triplet contributions:
\begin{eqnarray}
\Delta^{\rm sing}_\b{k}&\equiv&\frac{1}{2}(\Delta_{\b{k}\uparrow\downarrow}-\Delta_{\b{k}\downarrow\uparrow})\label{eq:sing}\\
\Delta^{x}_\b{k}&\equiv&\frac{1}{2}(\Delta_{\b{k}\downarrow\downarrow}-\Delta_{\b{k}\uparrow\uparrow})\label{eq:trip1}\\
\Delta^{y}_\b{k}&\equiv&\frac{-i}{2}(\Delta_{\b{k}\downarrow\downarrow}+\Delta_{\b{k}\uparrow\uparrow})\label{eq:trip2}\\
\Delta^{z}_\b{k}&\equiv&\frac{1}{2}(\Delta_{\b{k}\uparrow\downarrow}+\Delta_{\b{k}\downarrow\uparrow}).\label{eq:trip3}
\end{eqnarray}
Since $\Delta_{-\b{k}}=-\Delta_\b{k}^T$, and $g_{\b{k}}=g_{-\b{k}}$, the singlet gap equation only contains contributions from the part of the interaction that is parity-even.
\begin{eqnarray}
\Delta^{\rm sing}_\b{k}&=&-\frac{1}{2N}\sum_{\b{k}'}[(V_{\uparrow\downarrow\downarrow\uparrow}-V_{\downarrow\uparrow\downarrow\uparrow})\Delta_{\b{k}'\uparrow\downarrow}\nonumber\\
&&+(V_{\uparrow\downarrow\uparrow\downarrow}-V_{\downarrow\uparrow\uparrow\downarrow})\Delta_{\b{k}'\downarrow\uparrow}]g_{\b{k}'}\\
&=&-\frac{1}{N}\sum_{\b{k}'}[U+4V(s_\b{k}s_{\b{k}'}+d_\b{k}d_{\b{k}'})]\Delta^{\rm sing}_{\b{k}'}g_{\b{k}'}.\label{eq:Deltasing}\nonumber\\
\end{eqnarray}

Likewise, the triplet parts only contain contributions from the part of the interaction that is parity-odd. In fact, in this model all triplet components satisfy the same gap equation
\be
\Delta_\b{k}^\lambda=-\frac{1}{N}\sum_{\b{k}'}2V(\sin k_x\sin k_x'+\sin k_y\sin k_y')\Delta^\lambda_{\b{k}'}g_{\b{k}'},\label{eq:Deltatrip}
\ee
for $\lambda=x,y,z$.

The $\b{k}$-dependence of Eqs.~\eref{eq:Deltasing} and \eref{eq:Deltatrip} determines the appropriate ansatz for the gap functions:

\begin{eqnarray}
\Delta^{\rm sing}_\b{k}&=&\Delta_0+\Delta_ss_{\b{k}}+\Delta_dd_{\b{k}},\label{eq:Deltaansatz1}\\
\Delta^\lambda_{\b{k}}&=&\Delta^\lambda_x\sin k_x+\Delta^\lambda_y\sin k_y,\label{eq:Deltaansatz2}
\end{eqnarray}
and $|\Delta_\b{k}|^2=|\Delta_\b{k}^{\rm sing}|^2+\sum_\lambda|\Delta_\b{k}^{\lambda}|^2$.\\

These components satisfy the following gap equations:
\begin{eqnarray}
\Delta_0&=&-\frac{U}{N}\sum_{\b{k}'}g_{\b{k}'}(\Delta_0+\Delta_ss_{\b{k}'}+\Delta_dd_{\b{k}'}),\label{eq:d0}\\
\Delta_s&=&-\frac{4V}{N}\sum_{\b{k}'}s_{\b{k}'}g_{\b{k}'}(\Delta_0+\Delta_ss_{\b{k}'}+\Delta_dd_{\b{k}'}),\label{eq:ds}\\
\Delta_d&=&-\frac{4V}{N}\sum_{\b{k}'}d_{\b{k}'}g_{\b{k}'}(\Delta_0+\Delta_ss_{\b{k}'}+\Delta_dd_{\b{k}'}),\label{eq:dd}\\
\Delta^\lambda_x&=&-\frac{2V}{N}\sum_{\b{k}'}\sin k_x'(\Delta^\lambda_x\sin k_x'+\Delta^\lambda_y\sin k_y')g_{\b{k}'}\ \ \ \ \ \ \ \ \label{eq:dx}\\
\Delta^\lambda_y&=&-\frac{2V}{N}\sum_{\b{k}'}\sin k_y'(\Delta^\lambda_x\sin k_x'+\Delta^\lambda_y\sin k_y')g_{\b{k}'}.\label{eq:dy}
\end{eqnarray}

Note that the triplet channels are degenerate since we may write 
Eqs.~\eref{eq:dx} and \eref{eq:dy} 
as an eigenvector equation $\b{\Delta}^\lambda=A\b{\Delta}^\lambda$, where $A$ is a $2\times2$ matrix independent of $\lambda$ and $\b{\Delta}^\lambda=(\Delta^\lambda_x,\Delta^\lambda_x)^T$. In other words, for all three values of $\lambda$, 
$\b{\Delta}^\lambda$ is an eigenvector of $A$ with eigenvalue $1$, but this can only be the case if all the $\b{\Delta}^\lambda$ are equal to one another up to multiplicative constants which can be absorbed into the definition~\eref{eq:dx}. Specifically, if $\b{\Delta}^x/c_1=\b{\Delta}^y/c_2=\b{\Delta}^z/c_3$, then the rescaling 
$\b{\Delta}^\lambda\rightarrow\b{\Delta}^\lambda/\sqrt{|c_1|^2+|c_2|^2+|c_3|^2}$ allows us to write the spectrum (and the free energy) in terms of a single triplet component 
$\Delta^{\rm trip}_\b{k}\equiv\Delta_x\sin k_x+\Delta_y\sin k_y$ so that $E_{\b{k}}=\sqrt{\xi_\b{k}^2+|\Delta^{\rm sing}_\b{k}|^2+|\Delta^{\rm trip}_\b{k}|^2}$.

\section{Free Energy}\label{app:free}


Combining the quasi-particle energy density with the ground state energy density yields the internal energy density:
\begin{eqnarray}
u&=&\frac{2}{N} \sum_{\mathbf{k}} E_{\mathbf{k}} f(E_\b{k})+\frac{1}{N} \sum_{\mathbf{k}}\left(\xi_{\mathbf{k}}-E_{\mathbf{k}}\right)\nonumber\\
&&+\frac{1}{N}\sum_\b{k}\frac{\left|\Delta_{\mathbf{k}}\right|^{2}}{2 E_{\mathbf{k}}}\left(1-2 f\left(E_{\mathbf{k}}\right)\right)+\mu n.
\end{eqnarray}
The entropy density is that of a free Fermi gas:
\begin{eqnarray}
s&=&-\frac{2k_{B}}{N} \sum_{\mathbf{k}}[\left(1-f(E_\b{k})\right) \ln \left(1-f(E_{\b{k}})\right)\nonumber\\
&&+f(E_\b{k}) \ln f(E_\b{k})].
\end{eqnarray}
Combining these produces the mean-field free energy density $f_{\rm MF}=u-Ts$:
\begin{eqnarray}
f_{\rm MF}&=&\frac{1}{N}\sum_\b{k}\bigg(\xi_\b{k}-E_\b{k}+|\Delta_\b{k}|^2g_\b{k}\bigg)\nonumber\\
&&+\frac{2k_{\rm B}T}{N}\sum_\b{k}\ln(1-f(E_\b{k}))+\mu n\label{eq:fmf}.
\end{eqnarray}
$f_{\rm MF}$ gives the correct value of the free energy at the solutions to the gap equation, i.e. its critical points, but it is not the correct functional to minimize in order to obtain these critical points. That functional comes from the finite temperature variational theorem~\cite{feynman1955}
\be
f\leq f[\Delta_\b{k}]\equiv f_{\rm MF} +\langle H-H_{\rm MF}\rangle_{\rm MF},
\ee
where the mean-field Hamiltonian $H_{\rm MF}$ is given in Eq.~\eref{eq:HMF}. Of course, if we know the solutions to the gap equation, we can simply plug them into $f_{\rm MF}$ and compare the resulting free energies, but it will prove fruitful to work with the variational free energy $f[\Delta_\b{k}]$. Here, the mean-field expectation of an operator $X$ is given by
\be
\langle X\rangle_{\rm MF}\equiv\frac{1}{Z_{\rm{MF}}}\tr e^{-\beta H_{\rm{MF}}}X.
\ee
We must compute
\begin{eqnarray}\label{eq:trace}
&&\langle H-H_{\rm MF}\rangle_{\rm{MF}}\nonumber\\
&=&\frac{1}{2N}\sum_{\b{kk}'}\sum_{\alpha\beta\gamma\delta}V_{\alpha\beta\gamma\delta}(\b{k},\b{k}')\langle c^\dagger_{\b{k}\alpha} c^\dagger_{-\b{k}\beta}c_{-\b{k}'\gamma}c_{\b{k}'\delta}\rangle_{\rm{MF}}\nonumber\\
&&+\frac{1}{N}\sum_\b{k}|\Delta_\b{k}|^2g_\b{k}.\nonumber\\
\end{eqnarray}
The trace of the quartic term is readily evaluated in the eigenbasis of $H_{\rm MF}$. Using Wick's theorem and the fact that the occupation number distribution of the quasi-particles $\langle\gamma^\dagger_{\b{k}\alpha}\gamma_{\b{k}\alpha}\rangle$ is the Fermi function $f(E_\b{k})$, we obtain the identity
\be
\langle\gamma^\dagger_{\b{k}\alpha}\gamma_{\b{k}\beta}\gamma^\dagger_{\b{k}'\gamma}\gamma_{\b{k}'\delta}\rangle_{\rm{MF}}=\delta_{\alpha\beta}\delta_{\gamma\delta}f(E_\b{k})f(E_{\b{k}'}).
\ee
Using this identity and applying the Bogoliubov transformation \eref{eq:bogo_extend1}, \eref{eq:bogo_extend2}, we can evaluate the quartic term
\be\label{eq:quartic}
\langle c^\dagger_{\b{k}\alpha}c^\dagger_{-\b{k}\beta}c_{-\b{k}'\gamma}c_{\b{k}'\delta}\rangle_{\rm{MF}}=\Delta^*_{\b{k}\alpha\beta}\Delta_{\b{k}'\delta\gamma}g_\b{k}g_{\b{k}'}.
\ee
The free energy functional follows from this,
\begin{eqnarray}\label{eq:freeEfunc}
f[\Delta_\b{k}]&=&\frac{2k_BT}{N}\sum_{\b{k}}\ln(1-f(E_\b{k}))\nonumber\\
&&+\frac{1}{N^2}\sum_{\b{kk}'}V^{\rm sing}(\b{k},\b{k}')\Delta^{{\rm sing}*}_\b{k}\Delta^{\rm sing}_{\b{k}'}g_{\b{k}}g_{\b{k}'}\nonumber\\
&&+\frac{1}{N^2}\sum_{\b{kk}'}V^{\rm trip}(\b{k},\b{k}')\Delta^{{\rm trip}*}_\b{k}\Delta^{\rm trip}_{\b{k}'}g_{\b{k}}g_{\b{k}'}\nonumber\\
&&+\frac{1}{N}\sum_{\b{k}}(\xi_\b{k}-E_\b{k}+2|\Delta_{\b{k}}|^2g_\b{k})+\mu n,
\end{eqnarray}
where we have separated the interaction into its parity-even and odd parts:
\begin{eqnarray}
V^{\rm sing}(\b{k},\b{k}')&=&U+4V(s_\b{k}s_{\b{k}'}+d_\b{k}d_{\b{k}'}),\\
V^{\rm trip}(\b{k},\b{k}')&=&2V(\sin k_x\sin k_x'+\sin k_y\sin k_y').\nonumber\\
\end{eqnarray} 
Minimization of the free energy is an important problem that requires strict numerical control. Ensuring that one has obtained the \emph{global} minimum of this function is numerically difficult in a brute-force approach, which amounts to extensively searching for all local minima in the five-dimensional parameter space spanned by the $\Delta_i$ components. In this regard, we have made use of the theory of Morse functions
\cite{palais1963,morse1934,morse1934} to aid our numerical search for the global minimum, which allows us to effectively reduce the dimensionality of the parameter space one needs to explore from five to two in this model.

\section{Critical Points}\label{app:critical}

The gap equations for the extended Hubbard model \eref{eq:d0}-\eref{eq:dy} can be reproduced from the solutions to the functional minimization problem:
\be\label{eq:gapeqn2}
\frac{\delta f}{\delta\Delta^{\rm sing}_\b{q}}=0;\;\;\frac{\delta f}{\delta\Delta^\lambda_\b{q}}=0,
\ee
while the chemical potential is fixed by the number equation
\be
\frac{\partial f}{\partial \mu}=0,
\ee
which upon using \eref{eq:gapeqn2} becomes,
\be\label{eq:number}
n=1-\frac{2}{N}\sum_\b{k}\xi_\b{k}g_\b{k}.
\ee
The functional minimization of $f$ can be thought of as a minimization with respect to the $N$ terms of $\Delta_\b{k}$ (one for each value of $\b{k}$). However, we can alternatively think of this as a minimization over a lower-dimensional parameter space spanned by ($\Delta_0$, $\Delta_s$, $\Delta_d$, $\Delta_x^\lambda$, $\Delta_y^\lambda$), so that we need only solve
\be
\frac{\partial f}{\partial \Delta_i}=\frac{\partial f}{\partial \Delta^\lambda_i}=\frac{\partial f}{\partial \mu}=0\label{eq:gap_num}
\ee 
for the all the components $\Delta_i\in\{\Delta_0, \Delta_s, \Delta_d\}$ and $\Delta^\lambda_i\in\{\Delta_x^\lambda, \Delta_y^\lambda\}$. The first derivative is given by
\begin{eqnarray}\label{eq:Fgrad}
\frac{\partial f}{\partial\Delta_i}&=&\frac{2}{N}\sum_\b{k}x_\b{k}^iG_\b{k}\bigg[\Delta^{\rm sing*}_\b{k}+\frac{1}{N}\sum_{\b{k}'}V^{\rm sing}_{\b{kk}'}\Delta^{\rm sing*}_{\b{k}'}g_{\b{k}'}\bigg],\nonumber\\
\\
\frac{\partial f}{\partial\Delta^\lambda_i}&=&\frac{2}{N}\sum_\b{k}x_\b{k}^iG_\b{k}\bigg[\Delta^{\lambda*}_\b{k}+\frac{1}{N}\sum_{\b{k}'}V^{\rm trip}_{\b{kk}'}\Delta^{\rm \lambda*}_{\b{k}'}g_{\b{k}'}\bigg],\label{eq:Fgrad2}
\end{eqnarray}
where we have defined $G_\b{k}\equiv g_\b{k}+\frac{|\Delta_\b{k}|^2}{E_\b{k}}\frac{\partial g_\b{k}}{\partial E_\b{k}}$ and used the basis functions~\eqref{eq:basis1}-\eqref{eq:basislast}. Since $G_\b{k}$ is positive definite, these equations simply restate what we already know; the critical points of $f$ occur at the values of $\Delta_i$ that satisfy the gap equation. One conceptual advantage of this formulation is that we can now compute a simple Hessian in a low-dimensional parameter space. For now, let us focus on real singlet order parameters to better understand these critical points. In that case, the Hessian elements evaluated at the critical points are
\begin{eqnarray}\label{eq:curvature}
\mathcal{H}_{ij}&\equiv&\frac{\partial^2f}{\partial\Delta_i\partial\Delta_j}\bigg|_{\rm c.p.}\nonumber\\
&=&\frac{2}{N}\sum_{\b{k}}x^i_{\b{k}}G_{\b{k}}\bigg(x^j_{\b{k}}+\frac{1}{N}\sum_{\b{k}'}V^{\rm sing}_{\b{kk}'}G_{\b{k}'}x^j_{\b{k}'}\bigg),
\end{eqnarray}
where $G_\b{k}$ is evaluated at the solutions to the gap equation.

In this case the critical points have two symmetries apparent from the gap equation. First, there is inversion symmetry; if $(\Delta_0, \Delta_s, \Delta_d)$ is a critical point, then so is $(-\Delta_0, -\Delta_s, -\Delta_d)$. This is required by antisymmetry of the pair wavefunction. Second, there is mirror symmetry under reflection in the $\Delta_0$-$\Delta_s$ plane, since any solution $(\Delta_0, \Delta_s, \Delta_d)$ has a corresponding solution $(\Delta_0, \Delta_s, -\Delta_d)$ upon replacing $k_x\leftrightarrow k_y$, $k_x'\leftrightarrow k_y'$.  Moreover, the curvature \eref{eq:curvature} is a rank-2 tensor under these transformations. That is, under inversion,
\be
\mathcal{H}_{ij}\rightarrow \sum_{pq}\delta_{ip}\delta_{jq}\mathcal{H}_{pq},
\ee 
and under reflection in the $\Delta_0$-$\Delta_s$ plane it follows the Householder transformation~\cite{householder1958}
\be
\mathcal{H}_{ij}\rightarrow \sum_{pq}(\delta_{ip}-2\delta_{id}\delta_{pd})(\delta_{jq}-2\delta_{jd}\delta_{qd})\mathcal{H}_{pq}.
\ee 
This means that the gradient flow of $f$ near the critical points respects these symmetries, and the \emph{index} of any critical points related by these symmetries is the same. The index $\gamma_i$ of a critical point $i$ is defined as the number of orthogonal directions along which $i$ is a maxima of the corresponding function (in this case the free energy). The index plays a large role in the theory of Morse functions~\cite{palais1963}. A Morse function is a smooth real function that has no degenerate critical points. For our purposes, the free energy is a Morse function, since the curvature can only vanish on a set of measure zero in the $\{T,U,V,n\}$ parameter space. As a consequence, the free energy satisfies the following Morse condition (valid for functions $f(\{\Delta_i\})$ that increase without bound as each $|\Delta_i|$ goes to infinity):
\be\label{eq:Morse}
\sum_{i\in\{\rm critical\; points\}}(-1)^{\gamma_i}=\chi(M),
\ee
where $\chi(M)$ is the Euler characteristic of the domain manifold of $f$~\cite{morse1934}. We may choose $M$ to be the three-dimensional space $\{\Delta_0,\Delta_s,\Delta_d\}$, or any of the pure subsets $\{\Delta_0, \Delta_s\}$ or $\{\Delta_d\}$, because any extrema of $f$ on a pure subset is guaranteed to be a solution to the full gap equation. All of these cases correspond to flat manifolds with $\chi(M)=1$. The combination of symmetries of $f$ and the Morse condition restrict the possible minima of the free energy in this parameter space.
\subsection{Corollaries of the Morse condition}
Continuing with the case of a real singlet gap function, the simplest question to ask is what can happen at $T_c$ according to these restrictions. Above $T_c$, the system is in the normal state, there is one critical point, and on any of the above manifolds, Eq. \eref{eq:Morse} reads
\be
(-1)^0=1.
\ee
We immediately see that no mixed solution can emerge from this state. This is because mixed solutions are four-fold degenerate by inversion and mirror symmetry, so that the Morse condition reads
\be
(-1)^{\gamma_N}+4(-1)^{\gamma_m}=1,
\ee
denoting the normal index by $\gamma_N$ and the mixed index by $\gamma_m$. This equation has no integer solutions. Thus the condition expressed in Ref.~\cite{annett1990} that mixed solutions cannot emerge at $T_c$ is an immediate consequence of the symmetry of the gap equation.

We may also prove the uniqueness (modulo sign) of the pure $d$-wave solution. In fact, 
the following will hold for any single-parameter gap function, including the Hubbard model with only on-site attraction. Existing proofs of the uniqueness of the BCS gap solution are quite nontrivial~\cite{vansevenant1985}. Here we will see that it is a simple consequence of symmetry and the Morse condition.
\begin{proof}
For a single component gap $\Delta_\b{k}=\Delta x_{\b{k}}$, with a separable interaction $V_{\b{kk}'}=Vx_\b{k}x_{\b{k}'}$, the Hessian reads
\begin{eqnarray}
H&=&\frac{2}{N}\sum_\b{k}x_\b{k}^2G_\b{k}\bigg(1+\frac{V}{N}\sum_{\b{k}'}G_{\b{k}'}x_{\b{k}'}^2\bigg)\\
&=&\frac{2}{N}\sum_\b{k}x_\b{k}^2G_\b{k}\bigg(1+\frac{V}{N}\sum_{\b{k}'}x_{\b{k}'}^2\bigg[g_{\b{k}'}+\frac{|\Delta_{\b{k}'}|^2}{E_{\b{k}'}}\frac{\partial g_{\b{k}'}}{\partial E_{\b{k}'}}\bigg]\bigg).\nonumber\\
\end{eqnarray}
The single-component gap equation reads
\be
1=-\frac{V}{N}\sum_{\b{k}}x_{\b{k}}^2g_\b{k},
\ee
so that at a critical point
\be
H=\frac{2V}{N^2}\sum_{\b{kk}'}x_\b{k}^2G_\b{k}x_{\b{k}'}^2\frac{|\Delta_{\b{k}'}|^2}{E_{\b{k}'}}\frac{\partial g_{\b{k}'}}{\partial E_{\b{k}'}}.\nonumber\\
\ee
$\frac{\partial g_{\b{k}'}}{\partial E_{\b{k}'}}$ is negative definite, while $G_{\b{k}}$ is positive definite, so for any attractive interaction $V<0$, the curvature is positive and all solutions must be minima of the free energy. Since the solutions are symmetric under inversion $\Delta\rightarrow-\Delta$, they must come in pairs. For $n$ such pairs,  the Morse condition reads
\be
(-1)^{\gamma_N}+2n(-1)^0=1,
\ee 
whose only solutions are $\{\gamma_N=0, n=0\}$ and $\{\gamma_N=1, n=1\}$ \footnote{Since this is a one-dimensional problem, this last statement is just Rolle's theorem.}. Thus, the single-component BCS gap equation admits one solution modulo sign.
\end{proof}
Returning to the extended Hubbard model, this result applies to solutions that are pure $d$-wave (in this case, $x_\b{k}\rightarrow d_\b{k}$ and $V\rightarrow4V$ in the arguments above), guaranteeing their uniqueness as well as the instability of the normal state. At a critical $d$-wave temperature $T_c^d$, two $d$-wave critical points emerge from the normal state critical point as the temperature is lowered. In fact, this kind of bifurcation of critical points is generic.

\begin{figure}[t]
	\centering
	\includegraphics[width=0.92\columnwidth]{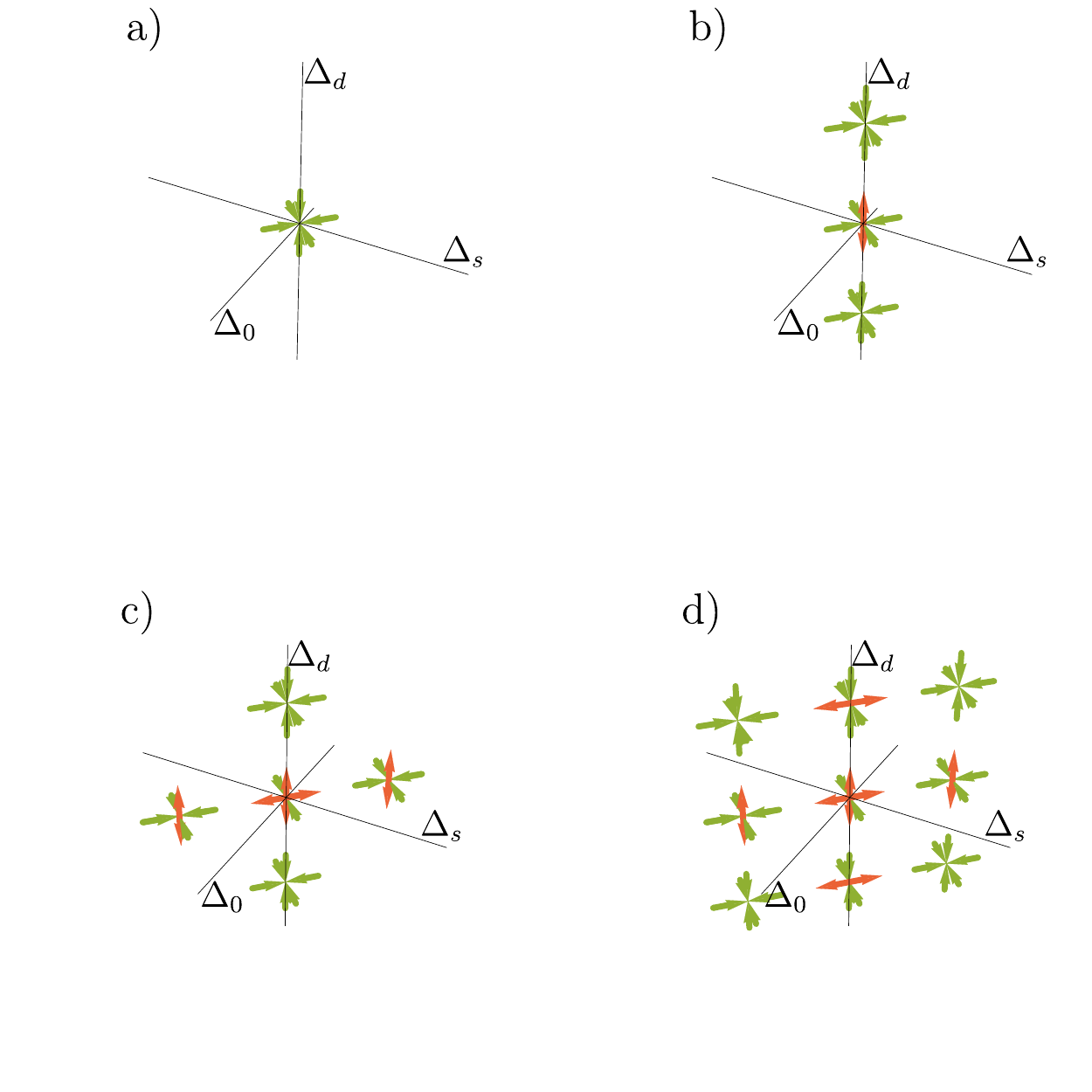}
\caption[Bifurcation of critical points]{Illustration of an evolution of solutions to the gap equation within the $\{\Delta_0, \Delta_s, \Delta_d\}$ subspace from high temperature (a), to low temperature (d). Inwards green arrows indicate minima of the free energy along those directions and outward red arrows indicate maxima along those directions. This example shows how the $s+id$ state is formed.} 
\label{fig:bifur}
\end{figure}

The solutions to the gap equation are continuous functions of $T$, so critical points cannot appear in pairs at arbitrary points in the parameter space but must grow from the normal state or an existing superconducting state as the temperature is lowered. The Morse condition then provides a conservation of indices. For example, a pure state with index $\gamma_p$ that grows from the normal state at a temperature $T_c^p$ is two-fold degenerate (barring pathological accidental degeneracies). This degeneracy comes from the inversion symmetry of the zeros of Eqs.~\eref{eq:Fgrad}, \eref{eq:Fgrad2} in the $\Delta_i$ parameter space. The corresponding critical points therefore must satisfy
\be
(-1)^{\gamma_{N_1}}=(-1)^{\gamma_{N_2}}+2(-1)^{\gamma_{p}},
\ee
where $\gamma_{N_1}$ is the index of the normal state at $T>T_c^p$, and $\gamma_{N_2}$ is the index of the normal state at $T<T_c^p$. The solution is
\be\label{eq:bifur}
\pm1=\mp1+2(\pm1).
\ee
So the normal state bifurcates into two pure states passing its index to the new solutions. Mixed states can grow out of pure states via the same bifurcation mechanism. 
An example of the evolution of critical points as the temperature is lowered is shown in Fig.~\ref{fig:bifur}.



\bibliography{SymmetryMixing}
\end{document}